\documentclass[aps, twocolumn, pra, showpacs, superscriptaddress]{revtex4}
\usepackage{amsfonts}
\usepackage{mathrsfs}
\usepackage{graphicx}
\usepackage{dcolumn}
\usepackage{bm}

\begin{document}

\title{Comparative study of quantum dynamics of a few bosons in a
one-dimensional split hard-wall trap: exact results versus
Bose-Hubbard-model approximations}
\author{Hongli Guo}
\affiliation{Institute of Physics, Chinese Academy of Sciences, Beijing 100190, China}
\author{Xiangguo Yin}
\affiliation{Institute of Physics, Chinese Academy of Sciences, Beijing 100190, China}
\author{Shu Chen}
\email{schen@aphy.iphy.ac.cn}
\affiliation{Institute of Physics,
Chinese Academy of Sciences, Beijing 100190, China}
\date{\today}

\begin{abstract}
We study the dynamical properties of a few bosons confined in an
one-dimensional split hard wall trap with the interaction strength
varying from the weakly to strongly repulsive regime. The system is
initially prepared in one side of the double well by setting the
barrier strength of the split trap to be infinity and then the
barrier strength is suddenly changed to a finite value. Both exact
diagonalization method and Bose-Hubbard model (BHM) approximation
are used to study the dynamical evolution of the initial system. The
exact results based on exact diagonaliztion verify the enhancement
of correlated tunneling in the strongly interacting regime.
Comparing results obtained by two different methods, we conclude
that one-band BHM approximation can well describe the dynamics in
the weakly interacting regime, but is not efficient to give
quantitatively consistent results in the strongly interacting
regime. Despite of the quantitative discrepancy, we validate that
the form of correlated tunneling gives an important contribution to
tunneling in the large interaction regime. To get a quantitative
description for the dynamics of bosons in the strongly interacting
regime, we find that a multi-band BHM approximation is necessary.
\end{abstract}


\maketitle



%\section{\label{sec:level1}First-level heading:\protect\\ The line
%break was forced \lowercase{via} \textbackslash\textbackslash}

{\label{sec:level1}}

\section{Introduction}

Bose-Einstein condensates in double-well potentials have attracted
much attention in the past decades. As a paradigm model for studying
the competition effect of quantum tunneling and interaction, the
double well systems have been widely studied from many aspects
\cite{PRA77063601,PRA593868,JPA381235,PRA78013604,Smerzi,PRA554318,RMP73307,Liujie,Raghavan,Mahmud}.
Due to the experimental progress in manipulating ultracold atomic
gases,
%such as interferometry \cite{Andrews}%
%, quantum information processing \cite{PRA73033605}
%quantum phase transition \cite{nature41539,PRA79033617}
%and quantum superposition state \cite{PRL87180402}.
%Due to the experimental progress in manipulating cold atoms,
both the trap potential and interaction between atoms can be
implemented with unprecedented tunability \cite{RMP}, and thus the
dynamics of many-body quantum states of interacting bosons can be
experimentally explored by loading the ultracold atoms in double
wells. For the atomic double-well system, the atom-atom interactions
in Bose-Einstein condensates have been found to play important roles
in the dynamics of the system. The competition between tunneling and
interaction leads to many rich and interesting effects, such as the
Josephson oscillation and self-trapping phenomena
\cite{Smerzi,PRA554318,RMP73307,Liujie,PRA593868,Raghavan,Mahmud}.
Moreover, novel correlated tunneling dynamics in interacting atomic
systems characterized by a small number of particles has recently
been observed experimentally \cite{Folling}, which attracted
particular attention in the study of the dynamics of few-atom
systems \cite{Zollner,Liang}.

So far most of theoretical works for double-well systems are based on the
two-mode approximation \cite%
{Smerzi,PRA554318,RMP73307,Liujie,Raghavan,PRA593868}.
%Within the two-mode approximation, the tunneling dynamics of BECs
%in a double-well potential can be understood by either semiclassical
%approach \cite{RMP73307}, mean-field theory \cite{Smerzi} or
%Fock-space WKB method \cite{PRA78023611}.
For atoms confined in a double well potential, if all of them are
prepared in one well initially, they will oscillate in the form of
Josephson oscillation for weak enough interaction, or they will stay
in one well (self-trapping) when the interaction is above a critical
value. Both the Josephson oscillation and self-trapping phenomena
can be understood within the two-mode approximation and have been
observed experimentally in cold atomic systems \cite{Albiez}.
However, if interactions are strong enough, the mean-field theory
and the two-mode approximation are not expected to be valid as
higher orbits are occupied. In the strongly interacting regime,
strong interactions between atoms may fundamentally alter the tunnel
dynamics and result in a correlated
tunneling, which was explored most recently in ultracold atoms \cite%
{Zollner,Liang}. Theoretically, the correlated paring tunneling was
studied by multi-configuration time-dependent Hartree method
\cite{Zollner,Sakmann} and also in the scheme of the extended
Bose-Hubbard model with an additional term of correlated pair tunneling \cite%
{Liang}. In order to understand the dynamics of double well systems
from weakly to strongly interacting regime in an unified scheme, in
this work we study the dynamical properties of a few bosons confined
in a one-dimensional (1D) split hard wall trap with the repulsive
interaction strength varying from zero to infinity. Experimentally,
the effective interaction strength can be tuned by using Feshbach
resonance or the confinement-induced resonance to the strongly
interacting Tonks-Girardeau (TG) limit \cite{Paredes,Weiss}, which
makes it possible to explore the novel dynamics even in the TG
limit.

In the TG limit, the bosonic systems exhibit the feature of
fermionization
\cite{Girardeau,Cederbaum,Hao1,Hao2,Zollner2,Deuretzbacher}. In the
strongly interacting regime, the mean field theory generally fails
to describe the properties of fermionization. In order to
characterize the crossover from weakly interacting condensation to
strongly interacting TG gas, some sophistical theoretical methods,
such as the exact diagonalization method
\cite{Deuretzbacher,Hao2,PRA78013604} and multi-orbital
self-consistent Hartree method \cite{Zollner2,Cederbaum} have been
applied to study the static few-boson systems. In this work, we
shall apply the exact diagonalization method to study the dynamical
problem in the 1D double-well system. The exact diagonalization
method can produce numerically exact results and allows us to give
an unified description for both the weakly and strongly interacting
regime. As a comparison, we also investigate the dynamics based on
the two-site Bose-Hubbard model by considering both the two-mode and
muti-mode approximations. Comparing the results obtained from
different methods, we conclude that one-band (two-mode) BHM
approximation is efficient to describe dynamics in small interaction
regime, but the multi-band BHM approximation is needed if we want to
describe the dynamics of bosons with large interaction
quantitatively. We also validate that the form of pair tunneling
gives an important contribution to tunneling in the large
interaction regime.

\section{Model and method}

We consider a few bosons with mass $m$ confined in an one-dimensional split
hard wall trap, which is described by the Hamiltonian ($\hbar =m=1$)
\begin{eqnarray}
\widehat{H} &=&\int \widehat{\psi }^{\dagger }(x) \left[ - \frac{1}{2}\frac{%
\partial ^{2}}{\partial x^{2}}+V(x)+\kappa \delta (x) \right] \widehat{\psi }%
(x) dx +  \nonumber \\
&& c\int \widehat{\psi }^{\dagger }(x)\widehat{\psi }^{\dagger }(x^{\prime
})\delta (x-x^{\prime })\widehat{\psi }(x)\widehat{\psi }(x^{\prime })
dxdx^{\prime }.  \label{Hamiltonian1}
\end{eqnarray}%
Here $V(x)$ is a hard wall trap which is zero in the region $(-a,a)$ and
infinite outside, $\kappa $ is a tunable parameter which describes the
strength of zero-ranged barrier at the center of the trap, and $c$ is the
interaction strength between particles determined by the effective \textrm{1D%
} s-wave scattering length. Here the double well is modeled by the
1D split hard-wall trap with a $\delta$-type barrier located at the
origin and the tunneling amplitude between the left and right wells
can be tuned by the barrier strength $\kappa $
\cite{PRA77063601,PRA78013604}. To study the tunneling dynamics, the
barrier strength $\kappa$ is initially set to be infinity and the
system is prepared
in the ground state of the left well. At time $t=0$, we suddenly change $%
\kappa$ to a finite value and study the dynamical evolution of the initially
prepared system.

\begin{figure}[tbp]
\includegraphics[height=8cm,width=8cm]{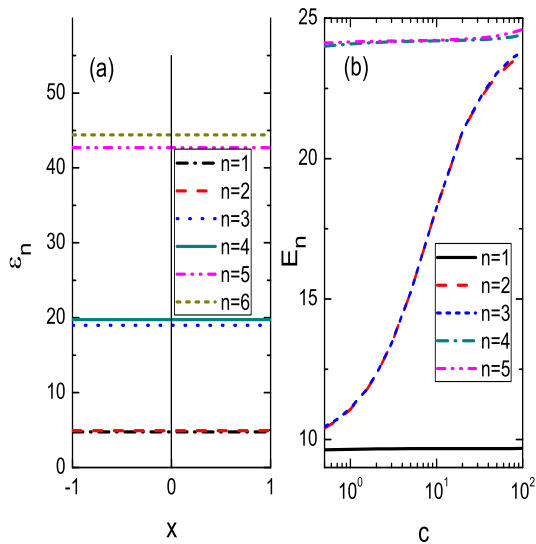}
\caption{(a)Single-particle energy levels for the split hard wall
trap with $\kappa=50$. (b) The eigenenergy of two interacting bosons
changing with the interaction strength $c$ in the split hard wall
trap with $\kappa=50$.} \label{f1}
\end{figure}

For $c=0$, the single particle stationary Schr\"{o}dinger equation
associated with the Hamiltonian (\ref{Hamiltonian1}) can be written as
\begin{equation}
\left[ -\frac{1}{2}\frac{\partial ^{2}}{\partial x^{2}}+V(x)+\kappa
\delta (x)\right] \varphi _{n}(x)=\epsilon _{n}\varphi _{n}(x),
\end{equation}%
where $\varphi _{n}(x)$ are the complete set of orthonormal
eigenfunctions and $\epsilon _{n}$ the corresponding eigenenergies.
Here $n=1,2,3,\cdots $ gives the ordering number of the
single-particle energies. According to the parity symmetry of the
eigenfunctions, the state $\varphi _{n}(x)$ is symmetric for odd $n$
($n=2i-1$) and antisymmetric for even $n$ ($n=2i$). The
single-particle energies are ordered alternatively corresponding to
symmetric and antisymmetric states. For convenience, we also
represent $\varphi _{2i-1}(x)=\varphi _{i,S}(x)$ and $\varphi
_{2i}(x)=\varphi _{i,A}(x)$ with the subscript $S$ ($A$) indicating
the symmetric (antisymmetric) function. The single-particle
antisymmetric eigenfunctions are $\varphi _{i,A}(x)=\frac{1}{\sqrt{a}}\sin (%
\frac{i\pi x}{a}),$ $i=1,2,3,...$ with their corresponding eigenenergies $%
\epsilon_{i,A}=\epsilon_{2i}=(i\pi /a)^{2}/2$, and the
single-particle symmetric eigenfunctions are $\varphi
_{i,S}(x)=C[\cos (px)-\frac{\kappa }{p}\sin (px)]\theta (-x)+C[\cos
(px)+\frac{\kappa }{p}\sin (px)]\theta (x)$ with their corresponding
eigenenergies $\epsilon_{i,S}=\epsilon_{2i-1}=p^{2}/2$, where the
wave vector $p$ is determined by transcend equation $p/\kappa +\tan
(pa)=0$ and $ C $ is the normalization constant. It is true that the
barrier only influences symmetric eigenfunctions, but does not
influence antisymmetric eigenfunctions for any barrier strength
$\kappa $. Further, the $(2i-1)$-th eigenenergy is close to the
$2i$-th eigenenergy ($i=1,2,3...$) gradually with
the increase of barrier strength and they become degenerate in the limit $%
\kappa \rightarrow \infty $. For simplicity, we set $a=1$ and
discuss a large barrier strength $\kappa=50$ in this paper. In this
case, single-particle eigenenergies in split hard wall trap are
$\epsilon _{1}=4.74341$, $\epsilon _{2}=\pi ^{2}/2$, $\epsilon
_{3}=18.9764$, $\epsilon _{4}=2\pi ^{2}$, $\epsilon
_{5}=42,7073,...$ respectively, see Fig.\ref{f1}(a). In contrast to
the small energy gap between the $(2i-1)$-th state and the $2i$-th
state, there is a relatively very large energy gap between the
$2i$-th state and $(2i+1)$-th state, {\it i.e.},
$\epsilon_{i,A}-\epsilon_{i,S} \ll \epsilon_{i+1,S}-\epsilon_{i,A}$.

Expanding field operators as $\widehat{\Psi }(x)=\sum_{n=1}^{\infty
}\varphi _{n}(x)a_{n}$, the many-body Hamiltonian
(\ref{Hamiltonian1}) takes the form
\begin{equation}
\widehat{H}=\sum_{n}\epsilon _{n}a_{n}^{\dag
}a_{n}+c\sum_{n,m,p,q}I_{nmpq}a_{n}^{\dag }a_{m}^{\dag }a_{p}a_{q},
\label{Hamiltonian2}
\end{equation}%
where $a_{n}^{\dag }(a_{n})$ is bosonic creation (annihilation)
operator for a particle in the single particle energy eigenstate
$\varphi _{n}$. The
interaction integral parameters $I_{nmpq}$ are calculated through $%
I_{nmpq}=\int_{-a}^{a}\varphi _{n}(x)\varphi _{m}(x)\varphi
_{p}(x)\varphi _{q}(x)dx$. The eigenstate of this Hamiltonian can be
obtained by numerical exact diagonalization in the subspace of the
energetically lowest eigen-states of a noninteracting many-particle
system \cite{Deuretzbacher,PRA78013604}.

When the barrier strength $\kappa $ is large, the split hard wall
trap can be considered as a double well. Similar to the case of
optical lattices, the local Wannier functions $W_{L}^{i}(x)$
$(W_{R}^{i}(x))$ at the left
(right) well with the energy band indices $i$ can be defined as $%
W_{L}^{i}(x)=1/\sqrt{2}(\varphi _{i,S}(x)+\varphi _{i,A}(x))$ and $%
W_{R}^{i}(x)=1/\sqrt{2}(\varphi _{i,S}(x)-\varphi _{i,A}(x))$. From the
symmetry of $\varphi _{i,S}(x)$ and $\varphi _{i,A}(x)$, one observes that $%
W_{L}^{i}(x)=W_{R}^{i}(-x)$. If we expand the bosonic field operator as, $%
\widehat{\psi
}(x)=\sum_{i}a_{i,L}W_{L}^{i}(x)+\sum_{i}a_{i,R}W_{R}^{i}(x)$ ,
where $a_{i,L\left( R\right) }$ is the bosonic annihilation operator
for a particle at left (right) well, the Hamiltonian can be written
as the form of two-site Bose Hubbard model
\begin{widetext}
\begin{eqnarray}
\widehat{H}=\sum_{i,j}(J_{LL}^{ij}a_{i,L}^{\dagger}a_{j,L}+J_{RR}^{ij}a_{i,R}^{\dagger}a_{j,R})
+\sum_{i,j}(J_{LR}^{ij}a_{i,L}^{\dagger}a_{j,R}+ J_{RL}^{ij}
a_{i,R}^{\dagger}a_{j,L})
+\sum_{i,j,k,l}\sum_{\alpha,\beta,\gamma,\delta}
U^{i,j,k,l}_{\alpha,\beta,\gamma,\delta}
a_{i,\alpha}^{\dagger}a_{j,\beta}^{\dagger}a_{k,\gamma}a_{l,\delta}
\label{Hamiltonian3}
\end{eqnarray}
\end{widetext}
where the integral $J_{\alpha \beta }^{ij}=\int_{-\infty
}^{\infty }dx(W_{\alpha }^{i}(x))^{\ast }H_{0}W_{\beta }^{j}(x)$, with $%
H_{0}=-\frac{1}{2}\frac{\partial ^{2}}{\partial x^{2}}+V(x)+\kappa \delta (x)
$ and the interaction integral $U_{\alpha ,\beta ,\gamma ,\delta
}^{i,j,k,l}=c\int dx(W_{\alpha }^{i}(x))^{\ast }(W_{\beta }^{j}(x))^{\ast
}W_{\gamma }^{k}(x)W_{\delta }^{l}(x)$. The subscripts $\alpha ,\beta
,\gamma ,\delta \in \{L,R\}$ are the well indices and the superscripts $%
i,j,k,l$$\in \{1,2,3,...\}$ are the energy band indices. Here we
note $J^{ij}_{RR}=J^{ij}_{LL}$ for the symmetric double well. The
Hamiltonian (\ref{Hamiltonian3}) can be divided into intraband and
interband parts, that is
\begin{equation}
\widehat{H}=\sum_{i}\widehat{H}_{i}+ \widehat{H}_{interband} .
\label{Hamiltonian4}
\end{equation}
The $i$-th intraband Hamiltonian can be written as
\begin{widetext}
\begin{eqnarray}
\widehat{H}_{i}&=&(\epsilon_{i,L} n_{i,L}+\epsilon_{i,R} n_{i,R})+
[J_{i}+2(n_{i,L}+n_{i,R}-1)J'_{i}]
(a_{i,L}^{\dagger}a_{i,R}+a_{i,R}^{\dagger}a_{i,L}) + \nonumber\\
& &
U_{0}^{i}[n_{i,L}(n_{i,L}-1)+n_{i,R}(n_{i,R}-1)]+U_{LR}^{i}(a_{i,L}^{\dagger}a_{i,L}^{\dagger}
a_{i,R} a_{i,R} +a_{i,R}^{\dagger} a_{i,R}^{\dagger} a_{i,L} a_{i,L}
+4 n_{i,L} n_{i,R}),\label{1bandBHM}
\end{eqnarray}
\end{widetext}
where $n_{i,L}=a_{i,L}^{\dagger}a_{i,L}$,
$n_{i,R}=a_{i,R}^{\dagger}a_{i,R}$, $ \epsilon_{i,L} = J_{LL}^{ii}$,
$ \epsilon_{i,R} = J_{RR}^{ii}$, $J_{i} = J_{LR}^{ii} = J_{RL}^{ii}$
is the intraband hopping energy between left and right wells,
$J'_{i} = U_{LLLR}^{iiii}=U_{RRRL}^{iiii}$, $U_{0}^{i} =
U_{LLLL}^{iiii}=U_{RRRR}^{iiii}$ is the on site interaction energy
and $U_{LR}^{i} \equiv U_{LLRR}^{iiii}$ is the intraband pair
hopping energy. It is easy to check that $\epsilon_{i,L}
=\epsilon_{i,R}=(\epsilon_{i,S} + \epsilon_{i,A} )/2 = \mu_i $ and
$J_{i}=(\epsilon_{i,S} - \epsilon_{i,A} )/2$. The interband
Hamiltonian reads
\begin{widetext}
\begin{equation}
\widehat{H}_{interband}=\sum_{i \neq
j}\sum_{\alpha,\beta,\gamma,\delta}
U^{i,j}_{\alpha,\beta,\gamma,\delta}
(a_{i,\alpha}^{\dagger}a_{j,\beta}^{\dagger}a_{i,\gamma}a_{j,\delta}+a_{i,\alpha}^{\dagger}a_{j,\beta}^{\dagger}a_{j,\gamma}a_{i,\delta})\nonumber\\
+\sum_{i,j,k,l}'\sum_{\alpha,\beta,\gamma,\delta}
U^{i,j,k,l}_{\alpha,\beta,\gamma,\delta}
a_{i,\alpha}^{\dagger}a_{j,\beta}^{\dagger}a_{k,\gamma}a_{l,\delta}
\end{equation}%
\end{widetext}
where $U^{i,j}_{\alpha,\beta,\gamma,\delta} =
U^{i,j,i,j}_{\alpha,\beta,\gamma,\delta}=
U^{i,j,j,i}_{\alpha,\beta,\gamma,\delta} $ and the summation $\sum'$
in the second part contains three-band terms where only three
different energy band indices exist and four-band terms with $i\neq
j\neq k \neq l$. Most interband interaction terms are very small
except for terms of $U_{LLLL}^{ij}$ and $U_{RRRR}^{ij}$ defined on
the same site and between the $i$th and $j$th bands. For the split
system with barrier strength $\kappa =50$, we have $ \epsilon_{i,L}
=\epsilon_{i,R} =4.83911$, $J_{1}=-0.0957$, $U_{0}^{1}=1.48446c$,
$J'_{1} =-0.00366c$, $U_{LR}^{1}=0.0003c$ and
$U_{LLLL}^{12}=0.98866c$, while the interaction strengths containing
three or four energy band indices are very small, for example
$U_{LLLL}^{1233}=-0.000969c$ and $U_{LLRR}^{1234}=-0.00107c$ .

Now we turn to consider the dynamical behavior for the $N$-boson
system. Initially (for $t<0$), the barrier strength $\kappa $ is set
to infinity and $N$ bosons are prepared in the ground state of the
left well. In this case, the split hard-wall trap actually reduces
to two separated hard-wall traps, so the eigenvalues and
eigenfunctions of a single particle in the left well are $(\pi
n/a)^{2}/2$ and $ \varphi _{n}(x)=-\sqrt{2/a}\sin (n\pi x/a)$, with
$n=1,2,3...$ respectively. Then at $t=0$, $\kappa$ is suddenly
changed to a finite value, for example $\kappa=50$ in the present
work, and the corresponding Hamiltonian is
$\widehat{H}_{f}=\widehat{H}(\kappa=50)$. After $\kappa $ is
changed, the time-dependant wave function is given by
\begin{equation}
|\Psi (t)\rangle =e^{-i\widehat{H}_{f}t}|\Psi (0)\rangle
=\sum_{n=1}^{\infty }C_{n}e^{-iE_{n}t}|\Phi _{n}\rangle ,
\label{Psit}
\end{equation}
in which weight coefficients are $ C_{n}=\langle \Phi _{n}|\Psi
(0)\rangle $ and they satisfy normalization condition
$\sum_{n=1}^{\infty }C_{n}^{2}=1$. Here $\Phi _{n}$ and $E_{n}$ are
the eigenstates and eigenvalues of $\widehat{H}_{f}$, respectively.
In order to see how the initial state trapped in the left trap
evolves, we shall use revival probability
\begin{eqnarray}
F(t) &=&|\langle \Psi (t)|\Psi (0)\rangle |^{2}  \nonumber \\
&=&1-4\sum_{n<m}C_{n}^{2}C_{m}^{2}\sin ^{2}[(E_{n}-E_{m})t/2],
\label{Fidelity}
\end{eqnarray}%
the reduced single-particle density matrix
\begin{equation}
\rho (x,x^{\prime },t)=\langle \Psi (t)|\widehat{\Psi }^{\dagger }(x)%
\widehat{\Psi }(x^{\prime })|\Psi (t)\rangle , \label{rhot}
\end{equation}
and pair correlation function
\begin{equation}
g^{(2)}(x_{1},x_{2},t)=\langle \Psi (t)|\widehat{\Psi }^{\dagger }(x_{1})%
\widehat{\Psi }^{\dagger }(x_{2})\widehat{\Psi }(x_{1})\widehat{\Psi }%
(x_{2})|\Psi (t)\rangle \label{g2}
\end{equation}
after time $t$ to describe the dynamics in split hard-wall trap
system.

\section{results and discussions}

Before studying the quantum dynamics of many-body systems, we first
recall the tunneling dynamics of a single atom. If there is only one
boson in this split hard wall trap, the initial state is just the
ground state of left well, that is $\Psi (0)=-\sqrt{2}\sin (\pi x)$
with energy $\pi ^{2}/2$. After the barrier strength $\kappa $
switches on to a finite but large strength, the weight coefficients
of ground and the
first excited state of the finial Hamiltonian $H_{f}$ are $C_{1}\approx \sqrt{%
2}/2$ and $C_{2}\approx \sqrt{2}/2$. At time $t$, the wavefunction
reads
\begin{eqnarray}
|\Psi (t)\rangle &\approx& \frac{\sqrt{2}}{2} e^{-i\epsilon_{1,S}
t}|\varphi_{1,S}\rangle + \frac{\sqrt{2}}{2} e^{-i\epsilon_{1,A}
t}|\varphi_{1,A}\rangle \nonumber \\
  &=&  e^{-i\mu_1
t} [ \cos(J_1 t) |W_{L}^1\rangle + i  \sin(J_1 t) |W_{R}^1\rangle ],
\label{single-psi}
\end{eqnarray}
where $\mu_1=(\epsilon_{1,S} +\epsilon_{1,A})/2$. It is obvious that
the boson stays in the left well with the probability of $\cos^2(J_1
t)$ whereas in the right well with the probability of $\sin^2(J_1
t)$. Consequently, the boson oscillates back and forth between two
wells with period $\tau =2\pi /(\epsilon_{1,A}-\epsilon _{1,S})=-
\pi/J_1$, which is influenced by the barrier strength $\kappa $
through controlling the energy gap between ground state and the
first excited state. Correspondingly, the fidelity $F(t) \approx
\cos^2(J_1 t)$ oscillates periodically between $1$ and $0$.

For a many-body system, no an analytical expression like
Eq.(\ref{single-psi}) is available. Nevertheless, when the atom
number is small, we can resort to the full exact diagonalization
method to calculate the energy spectrum and eigenstates via directly
diagonalizing the Hamiltonian (\ref{Hamiltonian2}). Consequently the
time-dependent wavefunction, revival probability, single-particle
density matrix and correlation function are straightforward to be
calculated via Eq.(\ref{Psit})-(\ref{g2}). For a continuum system,
we need truncate the set of single-particle basis functions to the
lowest $L$ orbitals (modes) and the basis dimension of a
$N$-particle system with $L$ orbitals (modes)  is given by $D=\left(
N+L-1\right) !/[N!\left( L-1\right) !]$. In general, one needs $L
\gg N$ and it is a formidable task to get the full spectrum of the
many-particle system as the particle number $N$ becomes large.
Therefore, despite the fact that the exact diagonalization method
can be applied to deal with the interacting boson systems in a
numerically exact way for all relevant interaction strengths, it
only restricts to small particle systems. When the interaction
strength is weak, the two-site Bose-Hubbard Hamiltonian under
single-band (two-mode) approximation is widely taken to be the model
system for the study of the dynamics of the double-well system. One
of advantages of the two-site Bose-Hubbard Hamiltonian
(\ref{Hamiltonian3}) is that every term in the Hamiltonian has a
straightforward physical meaning which can help us to understand the
physical consequence of different terms. Furthermore, in the scheme
of the two-site Bose-Hubbard model, the system is much more
tractable both analytically and numerically and a large system can
be studied. In the following, we shall first present exact numerical
results by exact diagonalization and then results based on the
two-site Bose-Hubbard model under two-mode (single-band) and
multi-mode (multi-band) approximations.

\subsection{Exact result by exact diagonalization}

We first consider the two-boson case. If two bosons are initially
prepared in the ground state of the left well as the initial state
of the system, which can be gotten by exact diagonalization method.
Through diagonalizing the second quantized initial Hamiltonian
$H_{in}$ in the Hilbert space spanned by the single-particle
eigenstates, we get the initial state $\Psi (0)$. Similarly the
eigenenergy and eigenvectors for $ H_{f}$ can also be obtained. As
an example, we plot the lowest five eigenenergies of two interacting
bosons in the split hard wall with $\kappa=50$ versus the
interaction strength $c$ in Fig. \ref{f1}(b).
%As to barrier strength $\kappa$ after time $t=0$, it need be large
%enough to make sure only some lower energystates are considered in
%$\Psi(t)$, the weight coefficients of high eigenstate is too small
\begin{figure}[tbp]
\includegraphics[height=10cm,width=\linewidth]{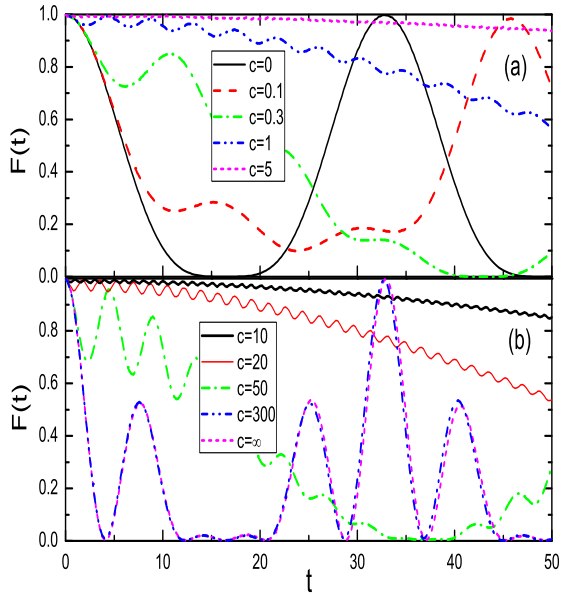}
\caption{The revival probability $F(t)$ for various $c$.} \label{f2}
\end{figure}

\begin{figure}[tbp]
\includegraphics[height=12cm,width=\linewidth]{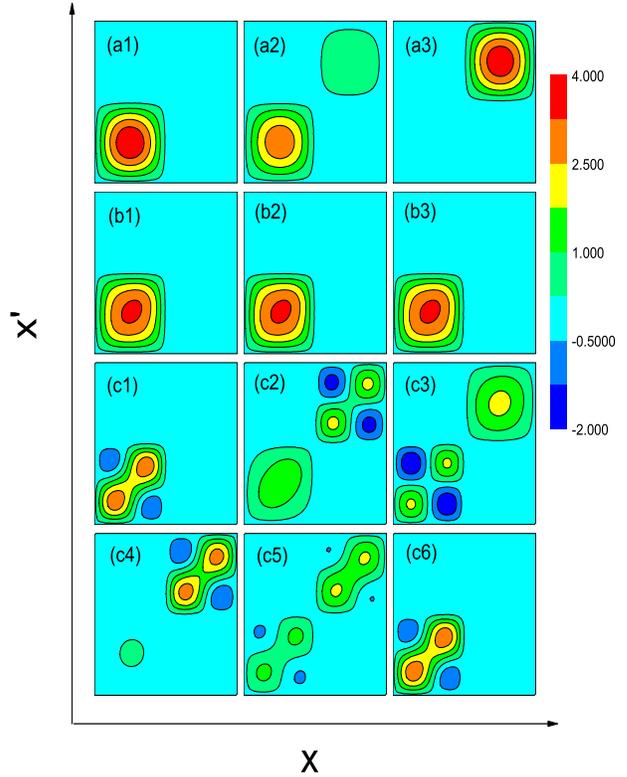}
\caption{(Color online) Reduced single-particle density matrix $\protect\rho %
(x,x^{\prime },t)$ of two interacting bosons for (a1-a3) $c=0,t=0,5,15$, (b1-b3) $%
c=5,t=0,5,15$, and (c1-c6) $c=\infty ,t=0,5,16,21,22.5,32$. Each
plot spans the range $-1<x,x^{\prime }<1$. } \label{f4}
\end{figure}

\begin{figure}[tbp]
\includegraphics[height=12cm,width=\linewidth]{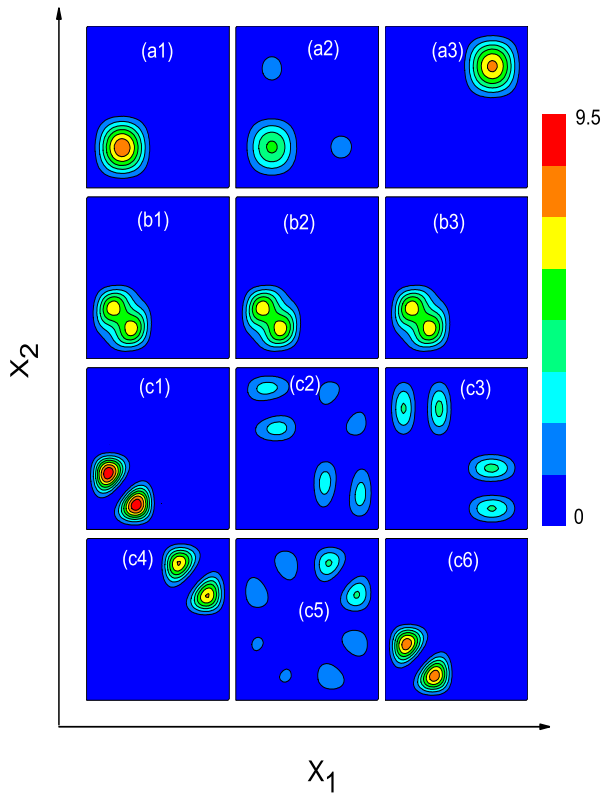}
\caption{(Color online) The pair correlation function $g^{(2)}(x_{1},x_{2},t)
$ for (a1-a3) $c=0,t=0,5,15$, (b1-b3) $c=5,t=0,5,15$, and (c1-c6) $c=\infty
,t=0,5,16,21,22.5,32$. Each plot spans the range $-1<x,x^{\prime }<1$. }
\label{f5}
\end{figure}

When the interaction is absent, bosons just oscillate back and forth
between two wells and return to their initial state after a Rabi
period $\tau $, which is the same as one boson's. For interacting
bosons, there will be many differences. The revival probability
$F(t)$ as a function of time $t$ is shown in Fig.\ref{f2} for
various $c$. When interaction strength $c$ is very small, the
revival probability $F(t)$ still displays the oscillating feature
and the system returns to their initial state with probability close
to $1$ after a longer period. The revival time becomes longer with
the increase of the interaction strength $c$. When $c$ reaches a
certain value ($c\sim 5$), $F(t)$ approaches $1$ with tiny
oscillations within a very large time scale, which is known as the
self trapping phenomena. In this regime, the tunneling to the right
well is dynamically suppressed and two bosons stay in the left well
stably. As the interaction strength $c$ increases further to the
stronger regime, $F(t)$ begins to decrease more quickly and two
bosons can tunnel to the right well again. In the limit of $c
\rightarrow \infty$, $F(t)$ approaches zero quickly and then
oscillates between $0$ and $0.521$, finally it approaches $1$ after
almost a Rabi period. To see clearly how the atoms tunnel between
the left and right wells, we display the corresponding
time-dependent reduced single particle density matrix $\rho
(x,x^{\prime },t)$ in Fig.\ref{f4} for several typical $c$. The
diagonal contribution $\rho (x,x^{\prime },t)$ along $x=x^{\prime }$
is just the single particle density distribution. In Fig.\ref{f5},
the pair correlation function $g^{(2)}(x_{1},x_{2},t)$ are also
displayed. The pair correlation function $g^{(2)}(x_{1},x_{2},t)$
shows the probability of finding one particle at point $x_{1}$ and
another particle at point $x_{2}$ in one measurement. As shown in
$(a1)$, $(a2)$ and $(a3)$ of Fig.\ref{f4} and Fig.\ref{f5}, the
non-interacting bosons can tunnel from left well to right well, and
then go back to left well, which forms a period of Rabi oscillation.
However, as shown in $(b1)$-$(b3)$ of Fig.\ref{f4} and Fig.\ref{f5},
the single particle density distribution and the pair correlation
function $g^{(2)}(x,x^{\prime },t)$ have no obvious change within a
Rabi period, and no oscillation between the left and right traps is
observed in this self trapping regime with $c=5$.
While in the fermionization limit, the oscillation phenomenon appears again in $%
(c1)-(c6)$. Both of two bosons tunnel to right well when $t=21$ (see
$(c4)$) and go back to the left well for $t=32$ (see $(c6)$). Our
results are consistent with results in \cite{Zollner} based on the
multi-configuration time-dependent Hartree method.

\begin{figure}[tbp]
\includegraphics[height=6cm,width=\linewidth]{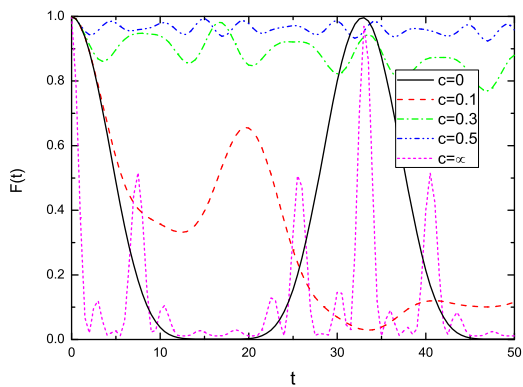}
\caption{The revival probability $F(t)$ of three particles system
changes with time $t$ with different interaction.} \label{f3}
\end{figure}
\

\begin{figure}[tbp]
\includegraphics[height=6cm,width=\linewidth]{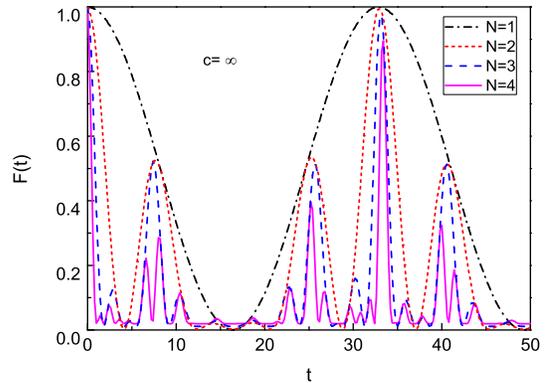}
\caption{The revival probability $F(t)$ of the $N$-boson system
changes with time $t$ with infinity interaction.} \label{f6}
\end{figure}

Next we consider the dynamics of systems with more bosons. The
dynamics of $N=3$ system is similar to the $ N=2$ case, except that
the system enters self-trapping regime earlier than $ N=2$. As shown
in Fig. \ref{f3}, when $c=0.5$, the system already displays the
feature of self trapping with the fidelity $F(t) \sim 1$ with tiny
oscillations within a very large time scale. Similarly, in the TG
limit, bosons can tunnel to the right well more easy and return to
the left well approximately after a Rabi period. The tunneling
dynamics of hard-core bosons is very similar to their correspondence
of free fermions \cite{Salasnich}.
%\begin{figure}
%\includegraphics[height=5cm,width=7cm]{c=0}
%\caption{The revival probability $F(t)$ of $n$ different particles
%changes with time $t$ with zero interaction.} \label{f4}
%\end{figure}
In Fig.\ref{f6}, we plot the revival probability $F(t)$ for systems
with $N=1$ to $N=4$ in the TG limit.  It is shown that there is an
obvious peak around the Rabi oscillation periods for various $N$,
which implies that the system can return to the left well with
probability close to $1$ after a Rabi period. One can understand
this from the Bose-Fermi mapping, i.e., bosons in the TG limit can
be mapped into a spinless free Fermi system \cite{Salasnich}. For
the case with $\kappa=50$, we can check that $J_i \approx i^2 J_1$
for $i = 1, 2, 3, 4$. The $N$ atoms initially occupy the $N$-lowest
single-particle levels of the left well, and roughly speaking, each
particle tunnels with the Rabi period $\tau_i = - \pi/J_i$. However,
when the particle number $N$ becomes large, the relations of $J_i
\approx i^2 J_1$ break down and even the dynamics in the TG limit
can be quite complex.

\subsection{BHM approximation}
Now we consider the two-site BHM described by the Hamiltonian
(\ref{Hamiltonian3}). If the interaction strength is much smaller
than the level spacing between the first band and the second band
defined as $\Delta=\mu_2 - \mu_1$, one may expect that the system
can be approximately described by the single-band BHM. Under the
one-band approximation, the Bose-Hubbard model is described by
Eq.(\ref{1bandBHM}) with $i=1$. We note that the pair hopping term
$U_{LR}^{1}$ in Eq.(\ref{1bandBHM}) is generally very small in
comparison with the on-site interaction, for example, in the present
work we have $U_0^{1} \approx 4948 U_{LR}^1 $. Therefore in many
previous works, the pair hopping term is omitted and a simplified
single-band BHM given by
\begin{eqnarray}
\widehat{H} &=& \mu_{1} (n_{1,L}+  n_{1,R})+ \tilde{J_{1}}
(a_{1,L}^{\dagger}a_{1,R}+a_{1,R}^{\dagger}a_{1,L}) + \nonumber \\
& & U_{0}^{1}[n_{1,L}(n_{1,L}-1)+n_{1,R}(n_{1,R}-1)],\label{SBHM}
\end{eqnarray}
has been widely used \cite{JPA381235,Ziegler,BHM,Wangli}. Here
$\tilde{J_{1}}=J_{1}+2(n_{1,L}+n_{1,R}-1)J'_{1}$. Since $U_{LR}^1
\ll U_{0}^1$ in the whole interacting regime, the pairing hopping
term is not expected to significantly change the static properties.
However, when the term $U_{LR}^1$ is comparable with the hopping
amplitude $J_1$, it may give significant contribution to the
dynamics, which has been emphasized in Ref.{\cite{Liang}}. In the
weakly interacting regime, the term of $J'_{1}$ is also usually
neglected as its revision to hopping energy can be attributed to
$J_{1}$. As we shall illustrate later, when the interaction strength
is very large, the contribution of $J'_{1}$ can not be neglected
since $J'_{1} \propto c$.
\begin{figure}[tbp]
\includegraphics[height=10cm,width=8cm]{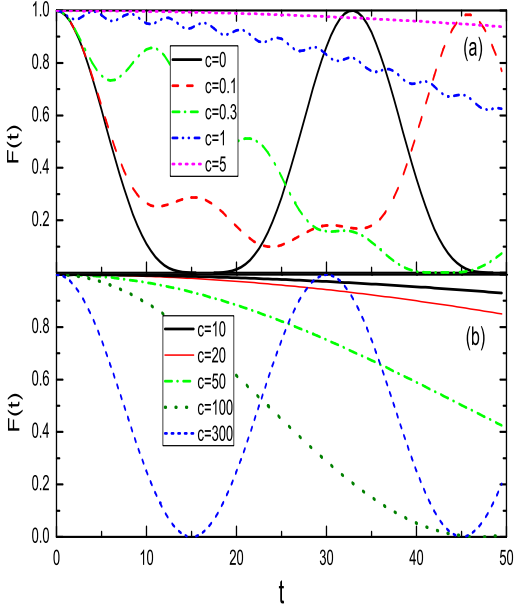}
\caption{The revival probability $F(t)$ for the two-boson system
changes with time $t$ under one-band BHM approximation.} \label{f7}
\end{figure}

\begin{figure}[tbp]
\includegraphics[height=7cm,width=\linewidth]{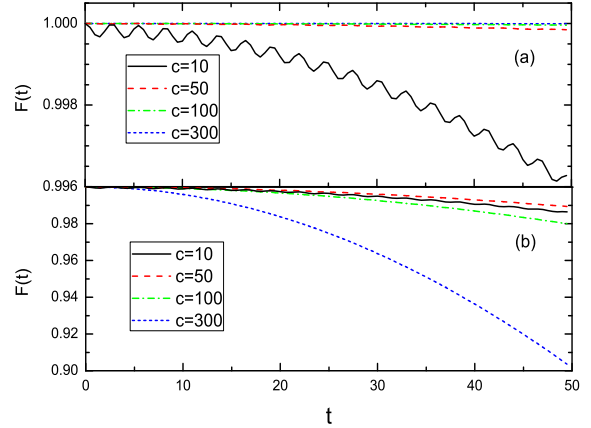}
\caption{The revival probability $F(t)$ for the two-boson system
changes with time $t$ under one-band BHM approximation with (a)
$U_{LR}^1=0$ and $J'_1 =0$. (b) $U_{LR}^1=0$. } \label{f8}
\end{figure}

By using the single-band BHM approximation, we study the tunneling
dynamics of the two-site BHM described by $\widehat{H}_1$ given by
Eq.(\ref{1bandBHM}) with all particles prepared in one site
initially. The revival probability $F(t)$ changing with time $t$
under one-band BHM approximation are shown in Fig.\ref{f7} for the
two-boson system. As shown in Fig.\ref{f7}a, when the interaction
strength is not very large, the single-band BHM gives quantitatively
consistent description of the dynamics in comparison with the exact
results by exact diagonalization (see Fig.\ref{f2}a). With further
increase in the interaction, as shown in Fig.\ref{f7}b, although the
single-band BHM with pairing hopping term can describe correctly the
enhancement of tunneling, it does not provide quantitatively
consistent results in comparison with results of exact
diagonalization in Fig.\ref{f2}b. In order to see clearly the effect
of the pair-tunneling term, we also study the dynamics governed by
the simplified BHM of Eq.(\ref{SBHM}) without the pair-tunneling
term. To see the effect of the term of $J'_1$, we consider both
cases for the Hamiltonian of Eq.(\ref{SBHM}) with or without the
term of $J'_1$. When the interaction strength is not so strong (for
example, $c < 5$), we find that the results are almost the same as
that presented in Fig\ref{f7}a. That means that the terms of pair
tunneling and $J'_1$ are not important when the interaction is weak.
However as shown in Fig.\ref{f8} (a) and (b), the dynamics in the
strongly interacting regime shows quite different behaviors if the
pair tunneling term is absent. Comparing Fig.\ref{f7}b and
Fig.\ref{f8}, we can conclude that the pair tunneling term of
$U_{LR}^{1}$ gives an important contribution to tunneling in the
large interaction regime. Comparing Fig.\ref{f8}a and Fig.\ref{f8}b,
we find that the term of $J'_1$ also plays an important role in
enhancing the tunneling.

\begin{figure}[tbp]
\includegraphics[height=6cm,width=\linewidth]{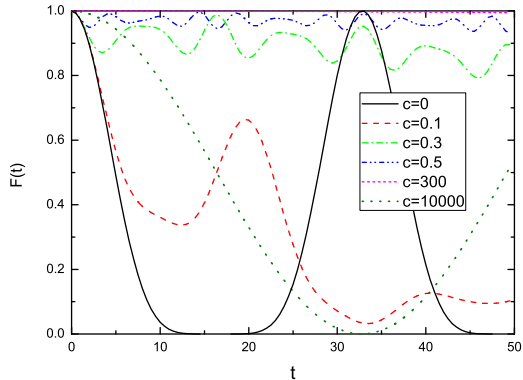}
\caption{The revival probability $F(t)$ for the three-boson system
changes with time $t$ under one-band BHM approximation.} \label{f9}
\end{figure}

The dynamics for the three-boson system is shown in Fig.\ref{f9}.
Comparing with the exact dynamical results in Fig.\ref{f3}, we find
that the one-band BHM approximation can describe the dynamics well
only when the interaction strength is small so that $U_{0}^{1} \ll
\Delta $. In contrast to the two-boson system, the pair-tunneling
term has less significant effect on the enhancement of the
tunneling. For very large $c$, although the system can tunnel to the
right well, it does not give quantitatively consistent results in
comparison with results of the numerical exact diagonalization.

\begin{figure}[tbp]
\includegraphics[height=6cm,width=\linewidth]{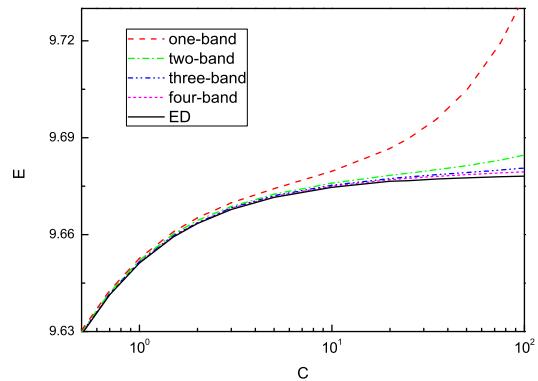}
\caption{The ground energy for the two-boson system obtained by
exact diagonalization method and  $i$-band BHM approximation with
$i=1,2,3,4$.} \label{f10}
\end{figure}

\begin{figure}[tbp]
\includegraphics[height=6cm,width=\linewidth]{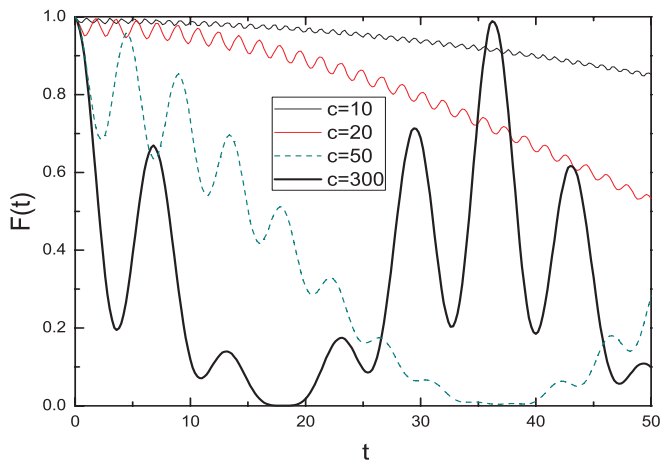}
\caption{The revival probability $F(t)$ changes with time $t$ under
multi-band BHM approximations, including a five-band BHM
approximation for $c=10$, a seven-band BHM approximation for $c=20$,
a ten-band BHM approximation for $c=50$ and  a eignteen-band BHM
approximation for $c=300$. } \label{f11}
\end{figure}

%\subsection{Multi-band BHM approximation}
From the above results, we know that one-band BHM approximation is
not enough to give a quantitatively description for the dynamics of
interacting bosons in the large interaction regime. To get better
results, we need keep more band levels and use the multi-band BHM
given by Eq.(\ref{Hamiltonian3}) in our calculation. In
Fig.\ref{f10}, we show the ground energy of the two bosons versus
the interaction strength using exact diagonalization method,
one-band, two-band, three-band and four-band BHM approximations,
respectively. It is shown that one-band BHM approximation can
describe the ground energy very well when the interaction $c<1$, and
two-band BHM approximation can describe well in the region $c<10$,
while three-band and four-band BHM approximations are efficient to
describe the ground energy of two bosons well even for $c=100$. For
the dynamics problem, in order to get a quantitatively consistent
results with the exact diagonaliztion results, we find that more
bands are needed in comparison with the static problem. In
Fig.\ref{f11}, we display the results of $F(t)$ for the two-particle
systems with various $c$ within the multi-band BHM approximation. As
shown in the figure, the result based on a five-band BHM
approximation for $c=10$ already quantitatively agrees with the
exact numerical result. For $c=50$, a ten-band BHM approximation is
required for a quantitatively consistent result. The result for
$c=300$ based on an eighteen-band BHM approximation is also given in
Fig.\ref{f11}. In comparison with Fig.2b and Fig.7b, we find that
there exists only a qualitative agreement with the exact
diagonaliztion result although it is much better than the result of
the single-band BHM approximation.

\section{Summary}

In summary, we have studied the dynamical properties of a few bosons
confined in an one-dimensional split hard wall trap by both the
exact diagonaliztion method and the approximate method based on the
two-site Bose-Hubbard model. The system is initially prepared in the
left well of the trap by setting the barrier strength of the split
hard wall trap  to infinity,  and then it is suddenly changed to a
finite value. With the increase in the interaction strength of
bosons, the system displays the Josephson-like oscillations, self
trapping and correlated tunneling in turn. Comparing results
obtained by two different methods, we conclude that the one-band BHM
approximation can quantitatively describe the dynamics in the weakly
interacting regime, but the multi-band BHM approximation is needed
if we want to describe the dynamics of bosons with large interaction
quantitatively. We also validate that the form of correlated
tunneling gives an important contribution to the tunneling dynamics
in the large interaction regime.

%\newpage
%\bibliographystyle{plain}
%\bibliography{apssamp}

\end{document}